\documentclass[twocolumn,showpacs,aps,prl,superscriptaddress]{revtex4}

\usepackage{graphicx}
\usepackage{dcolumn}
\usepackage{amsmath}
\usepackage{epsfig}

\input pubboard/babarsym
\input Definitions

\newcommand{\BABARPubYear}    {04}
\newcommand{\BABARPubNumber}  {017}

\newcommand{\SLACPubNumber}   {10498}
\newcommand{\LANLNumber}      {xxxxxx}

\def\figurebox#1#2#3{%
    \def\arg{#3}%
    \ifx\arg\empty
    {\hfill\vbox{\hsize#2\hrule\hbox to #2{\vrule\hfill\vbox to #1{\hsize#2\vfill}\vrule}\hrule}\hfill}%
    \else
    {\hfill\epsfbox{#3}\hfill}%
    \fi}

\begin{document}

\preprint{\babar-PUB-\BABARPubYear/\BABARPubNumber} 
\preprint{SLAC-PUB-\SLACPubNumber} 

\begin{flushleft}
\babar-PUB-\BABARPubYear/\BABARPubNumber\\
SLAC-PUB-\SLACPubNumber\\
hep-ex/\LANLNumber\\[10mm]
\end{flushleft}

\title{
{\large \bf \boldmath
Measurements of the Branching Fraction and 
{\em CP}-Violation Asymmetries 
in \boldmath$B^0\to\ffzeroKs$} 
}
%
\author{B.~Aubert}
\author{R.~Barate}
\author{D.~Boutigny}
\author{F.~Couderc}
\author{J.-M.~Gaillard}
\author{A.~Hicheur}
\author{Y.~Karyotakis}
\author{J.~P.~Lees}
\author{V.~Tisserand}
\author{A.~Zghiche}
\affiliation{Laboratoire de Physique des Particules, F-74941 Annecy-le-Vieux, France }
% ...
\collaboration{The \babar\ Collaboration}
\noaffiliation

\date{\today}

\begin{abstract}
We present measurements of the branching fraction and 
\CP-violating asymmetries in the decay $B^0\to\ffzeroKs$.
The results are obtained from a data sample of $123\times10^6$ 
$\FourS \to B\Bbar$ decays.
From a time-dependent maximum likelihood fit we measure the 
branching fraction 
${\cal B}(\Bz\to\fzero(\to\pi^+\pi^-)K^0)=(6.0\pm0.9\pm0.6\pm1.2)\tmsix$, 
the mixing-induced \CP\ violation parameter 
$S=-1.62^{+0.56}_{-0.51}  \pm 0.09\pm 0.04$
and the direct \CP\ violation parameter 
$C=0.27\pm 0.36\pm 0.10 \pm 0.07 $, 
where the first errors are statistical, the second systematic
and the third due to model uncertainties.
We measure the 
$\fzero$ mass and width to be 
$m_{\fzero}=(980.6 \pm 4.1 \pm 0.5\pm 4.0)\mevcc$ and 
$\Gamma_{\fzero}=(43^{+12}_{-9} \pm 3 \pm 9 )\mevcc$, 
respectively.

\end{abstract}


\maketitle

In the Standard Model (SM), \CP violation arises from a single
phase in the three-generation Cabibbo-Kobayashi-Maskawa 
quark-mixing matrix~\cite{CKM}. Possible indications of physics
beyond the SM may be observed in the time-dependent 
\CP asymmetries of  $B$ decays dominated by penguin-type diagrams
to states such as $\phi K^0$,
$\eta^\prime K^0$, $K^+K^-K^0$, and $\fzero K^0$ \cite{NewPhys}.
Neglecting CKM-suppressed amplitudes,
these decays carry  the same weak phase as the decay 
$\Bz\to\jpsi K^0$ \cite{phases}. As a consequence, their mixing-induced 
\CP-violation parameter is expected to be 
$-\eta_f\times\stwob = -\eta_f\times0.74\pm0.05$ \cite{HFAG} 
in the SM, where $\eta_f$ is the \CP eigenvalue of the final state 
$f$, which is +1 for $\ffzeroKs$.  
There is no direct \CP violation expected
in these decays since they are dominated by a single amplitude in the SM.
Due to the large virtual mass scales occurring in the penguin loops,
additional diagrams with non-SM heavy particles in the loops and new
\CP-violating phases may contribute. Measurements 
of \CP violation in these channels and their comparisons with the SM 
expectation are therefore sensitive probes for physics beyond the SM.

In this Letter we present a measurement of the branching fraction 
and \CP-violating asymmetries in the penguin-dominated decay  
$B^0\rar\fzeros\KS$ \cite{f0ref} from a time-dependent
 maximum likelihood analysis.
We also measure the  mass and width of the $\fzeros$ resonance. We restrict the analysis to the region of the 
$\pi^+ \pi^- \KS$ Dalitz plot that is dominated by the $\fzeros$ and
  we  refer to this as the quasi-two-body (Q2B) approach.  
Effects due to the interference between the $\fzeros$ 
and the other resonances in the Dalitz plot are taken as 
systematic uncertainties.  

The data we use in this analysis were accumulated with the \babar\ 
detector~\cite{bib:babarNim} at the \pep2 asymmetric-energy $e^+e^-$ 
storage ring at SLAC. The data sample consists of  an integrated 
luminosity of $111.2\invfb$  collected at the \FourS resonance
 (``on-resonance'') corresponding to $(122.6\pm0.7)\times10^{6}$ 
$\B\Bbar$ pairs,
and $11.8\invfb$ collected about $40\mev$ below 
the~\FourS (``off-resonance''). In Ref.~\cite{bib:babarNim} we describe 
the silicon vertex tracker and drift chamber used for track and vertex
reconstruction, and the \v{C}erenkov detector (DIRC), the electromagnetic 
calorimeter (EMC), and the instrumented flux return (IFR) used for particle 
identification.

If we denote by  $\deltat$  the difference between the proper 
times of the decay of  the fully reconstructed $B^0\rar f_0\KS$ ($\Bz_{\rm rec}$)
and the decay  of the  other meson ($\Bz_{\rm tag}$),  the time-dependent decay
rate $f_{{\rm Q_{tag}}}$ is given by 
\beqn
\label{eq:theTime}
f_{{\rm Q_{tag}}}(\deltat) & = & 
        \frac{e^{-\left|\deltat\right|/\tau}}{4\tau}
        \bigg( 1 + {\rm Q_{tag}} S\sin(\deltamd\deltat) \nonumber \\
  &&	 -\;{\rm Q_{tag}} C\cos(\deltamd\deltat)
        \bigg)\,,
\eeqn
where   $Q_{\rm tag}= 1(-1)$ when the tagging meson $\Bz_{\rm tag}$
is a $\Bz(\Bzb)$, $\tau$ is the mean \Bz lifetime, and $\deltamd$ is 
the $\BzBzb$ oscillation frequency corresponding to the mass difference. 
 The parameter $S$ 
is non-zero if there is  mixing-induced \CP violation while a non-zero value
for $C$ would indicate direct \CP violation.

We reconstruct $B^0\rar f_0 (\rar\pi^+\pi^-) \KS$ candidates from combinations 
of two  tracks and a $\KS$~decaying to $\pi^+\pi^-$.
For the $\pi^+\pi^-$ pair  from the $\fzeros$ candidate, 
we use information  from the tracking system, EMC, and DIRC to 
remove tracks consistent with electron, kaon, and proton hypotheses.  
In addition, we require at least one track to have a          
signature in the IFR that is inconsistent with the muon hypothesis.
The mass of the 
$\fzeros$ candidate must satisfy $0.86<m(\pi^+\pi^-)<1.10\gevcc$.  To reduce
combinatorial background from low energy pions, we 
require $|\cos{\theta(\pi^+)}|<0.9$, where  $\theta(\pi^+)$ 
is the angle between the positive pion in  the $\fzeros$  rest frame
 and the $\fzeros$ flight direction in the laboratory frame.  
The $\KS$ candidate is required to have a mass within $10\mevcc$ of 
the nominal $K^0$ mass \cite{PDG2002} and a decay vertex separated from the $B^0$ decay vertex
 by at least five standard deviations. In addition, the cosine 
of the angle between the $\KS$ flight direction and the vector between 
the $\fzeros$ and the $\KS$ vertices must be greater than 0.99.

Two kinematic variables are used to discriminate between signal-$B$ 
decays and combinatorial background. One variable is the difference 
$\de$ between the measured center-of-mass (CM) energy of the $B$~candidate and $\sqrt{s}/2$, where $\sqrt{s}$ is the CM energy. The other variable is the 
beam-energy-substituted mass 
$\mes\equiv\sqrt{(s/2+{\mathbf {p}}_i\cdot{\mathbf{p}}_B)^2/E_i^2-{\mathbf {p}}_B^2},$
where the $B$ momentum ${\mathbf {p}}_B$ and the four-momentum of the 
initial state ($E_i$, ${\mathbf {p}}_i$) are defined in the laboratory 
frame. We require $5.23 < \mes <5.29\gevcc$ and $|\de|<0.1\gev$.

Continuum $e^+e^-\to q\bar{q}$ ($q = u,d,s,c$) events are the dominant 
background.  To enhance discrimination between signal and continuum, we 
use an neural network (NN) to combine four  variables: the 
cosine of the angle between the $B^0_{\rm rec}$ direction 
and the beam axis in the CM,
the cosine of the angle between the thrust axis of the
$B^0_{\rm rec}$  candidate
and the beam axis, and the zeroth and second angular moments $L_{0,2}$
of the energy flow about the $B^0_{\rm rec}$ thrust axis.  The moments
are defined by $L_j=\sum_i p_i \times |\cos{\theta_i}|^j$ where $\theta_i$
is the angle with respect to the $B^0_{\rm rec}$ thrust axis of the track
or neutral cluster $i$ and $p_i$ is its momentum.  The sum excludes the 
$B^0_{\rm rec}$ candidate.
The NN is trained  with off-resonance data and
simulated signal events. The final sample of signal candidates 
is selected with a cut on the NN output that retains 
$\sim 97\%$ ($52\%$) of the signal (continuum).

The signal efficiency determined from Monte Carlo (MC) 
simulation is $(37.2\pm3.1)\%$.  MC simulation shows that  $4.7\%$ of the 
selected signal events are misreconstructed,  
 mostly due to combinatorial
background from low-momentum tracks used to form the $\fzeros$ candidate.
In total, 7,556 on-resonance  events pass all selection criteria.

We use  MC-simulated events to study the background from other $B$
decays. The charmless decay modes are grouped into eight classes with similar 
kinematic and topological properties. The modes that decay to the $\pi^+\pi^-\KS$ 
final state are of particular importance since they have signal-like 
$\de$ and $\mes$ distributions and their decay amplitudes interfere with 
the $\fzeros\KS$ decay amplitude.  Among these modes are $\rho^0\KS$, 
$\fzerop\KS$, $\ftwo\KS$, $K^{*+}\pi^-$ (including other kaon 
resonances decaying to $\KS\pi^+$), and non-resonant $\pi^+\pi^-\KS$ 
decays.  The inclusive charmless $\pi^+\pi^-\KS$ branching fraction 
$(23.4\pm3.3)\tmsix$~\cite{HFAG}, together with the available 
exclusive measurements~\cite{HFAG},  are used to infer upper limits on the
branching fractions of these decays. 
 Along with selection efficiencies
 obtained from MC, these branching fractions are used to estimate the expected
background.
 The charmed decays  
$B^0\rar D^-\pi^+\rar\KS\pi^-\pi^+$ and $\B^+\rar\Dzb\pi^+\rar\KS\pi^0\pi^+$  
contribute significantly to the selected data sample.  Each of these modes 
is  treated as a separate class. Two additional classes account for 
the remaining neutral and charged $b \to c$ decays.
In the selected data sample we expect $45\pm15$ charmless and $128\pm74$ 
$b \to c$ events.

The time difference $\deltat$ is obtained from the measured distance between 
the $z$ positions (along the beam direction) of the $\Bz_{\rm rec}$ and 
$\Bz_{\rm tag}$ decay vertices, and the boost $\beta\gamma=0.56$ of 
the \epem\  system~\cite{bib:BabarS2b,bib:BabarSin2alpha}. 
To determine the flavor of the $\Bz_{\rm tag}$ 
we use the tagging algorithm of Ref.~\cite{bib:BabarS2b}.
This produces four mutually exclusive tagging categories. We also 
retain untagged events in a fifth category to improve the efficiency 
of the signal selection.

We use an unbinned extended maximum likelihood fit to extract
the $\fzeros\KS$ event yield, the \CP parameters defined 
in Eq.~(\ref{eq:theTime}), and the $\fzeros$ resonance parameters. 
The likelihood function for the $N_\cat$ candidates tagged in category $k$ is
\begin{equation}
\label{eq:pdfsum}
{\cal L}_k = e^{-N^{\prime}_\cat}\!\prod_{i=1}^{N_\cat}
		\bigg\{ N_{S} 
	\epsilon_\cat {\cal P}_{i,\cat}^{S}
	+ N_{C,\cat} {\cal P}_{i,\cat}^{C} 
 	+ \sum_{j=1}^{n_B} N_{B,j} \epsilon_{j,\cat}{\cal P}^{\B}_{ij, \cat}
	\bigg\}
\end{equation}
where $N^{\prime}_\cat$ is the sum of the signal, continuum 
and the $n_B$ $B$-background yields tagged in category $\cat$,
 $N_S$ is the number of 
$\fzeros\KS$ signal events in the sample, $\epsilon_\cat$ is the 
fraction of signal events tagged in category $\cat$, $N_{C,\cat}$ 
is the number of continuum background events  that are tagged in 
category~$\cat$, and $N_{B,j}\epsilon_{j,\cat}$ is the number of
$B$-background events of class $j$ that are tagged in category~$\cat$.
The \B-background event yields are fixed parameters, with the exception 
of the $D^-\pi^+$ yield.  Since $\Bz\rar D^-\pi^+$ events have a 
characteristic distribution in $\cos{\theta(\pi^+)}$, well 
separated from continuum and $\fzeros\KS$ events, the $D^-\pi^+$ is
 free to vary in the fit along with the signal and continuum yields.
The total likelihood 
${\cal L}$ is the product of the likelihoods for each tagging category.

The probability density functions (PDFs)  ${\cal P}_{\cat}^{S}$,  
${\cal P}_{\cat}^{C}$ and ${\cal P}^{\B}_{j, \cat}$, for signal,
continuum background and $B$-background class $j$, respectively,
are the products of the PDFs of six discriminating variables.
The signal PDF is thus given by
${\cal P}_\cat^{S} = 
	{\cal P}^{S}(\mes)\cdot 
	{\cal P}^{S}(\de) \cdot
 	{\cal P}_\cat^{S}(\NN) \cdot 
	{\cal P}^{S}(|\cos{\theta(\pi^+)|}) \cdot 
	{\cal P}^{S}(m(\pi^+\pi^-)) 
	\cdot {\cal P}_\cat^{S}(\deltat)$, 
where ${\cal P}_\cat^{S}(\deltat)$ contains the time-dependent 
\CP parameters defined in Eq.~(\ref{eq:theTime}), diluted by the 
effects of mis-tagging and the~$\deltat$ resolution. 


The signal PDFs are decomposed into distinct distributions for correctly reconstructed and mis-reconstructed signal events.
The fractions of mis-reconstructed signal events per tagging 
category are estimated by MC simulation.  The $\mes$, $\de$, 
\NN, $|\cos{\theta(\pi^+)}|$, and $m(\pi^+\pi^-)$  PDFs 
for signal and $B$ background are taken from the simulation except for 
the means of the signal Gaussian PDFs for $\mes$ and $\de$, and the mass 
and width of the $\fzeros$, which are free to vary in the fit.  We use 
a relativistic Breit-Wigner function to parameterize the $\fzeros$
resonance.  
The $\deltat$-resolution function for signal and $B$-background 
events is a sum of three Gaussian distributions, with parameters 
determined by a fit to fully reconstructed $\Bz$ decays~\cite{bib:BabarS2b}.
The continuum $\deltat$ distribution is parameterized as the sum of 
three Gaussian distributions  with two distinct means and three distinct 
widths,  which are scaled by the $\dt$ per-event error. 
For the $B$-background  modes that are \CP eigenstates, the parameters 
$C$ and $S$ are fixed to 0 and $\pm \sin{2\beta}$, respectively, 
depending on their \CP eigenvalues. For continuum, four tag asymmetries 
and the five yields $N_{C,\cat}$  are free. 
The signal yield, $S$, $C$, and the $\fzeros$ mass and width are 
among the 41 parameters that are free to vary in the fit.  The majority of the 
free parameters are used to describe the shape of the continuum background.


\begin{table}
\begin{center}
\caption{ \label{tab:systematics}
        Summary of systematic uncertainties.  The uncertainties on $m_{\fzeros}$ and $\Gamma_{\fzeros}$ are in units of $\mevcc$.}
\small
\begin{tabular}{lccccc} \hline
&&& \\[-0.3cm]
  &&&  ${\cal B}$ & &\\
\rs{Error Source} &  \rs{$\Sfzks$} &  \rs{$\Cfzks$}  
	&$\times 10^{-6}$&\rs{$m_{\fzeros} $}&\rs{$\Gamma_{\fzeros}$}\\
\hline
&&& \\[-0.3cm]
Fitting Procedure           & 0.06 ~~  & 0.07 ~~ & ~~ 0.26 ~~ & 0.5 ~~ & 1.0 \\
$B$-background              & 0.05 ~~  & 0.05 ~~ & ~~ 0.30 ~~ & 0.2 ~~ & 3.0 \\
$\dt$ Model                 & 0.03 ~~  & 0.02 ~~ & ~~ 0.03 ~~ & 0.0 ~~ & 0.1 \\
Tagging Fraction            & 0.04 ~~  & 0.02 ~~ & ~~ 0.01 ~~ & 0.0 ~~ & 0.2 \\
Signal Model                & 0.02 ~~  & 0.01 ~~ & ~~ 0.03 ~~ & 0.1 ~~ & 0.2\\
DCS Decays                  & 0.01 ~~  & 0.04 ~~ & ~~ 0.00 ~~ & 0.0 ~~ & 0.0\\
$\dmd$ and $\tau$       & 0.01 ~~ & 0.00 ~~  & ~~ 0.01 ~~  & 0.0 ~~ & 0.1  \\
\hline
Sub-total                       & 0.09 ~~  & 0.10 ~~ & ~~ 0.40 ~~ & 0.5 ~~ & 3.2\\
Q2B Approximation           & 0.04  ~~ & 0.07 ~~& ~~ 1.21 ~~& 4.0 ~~ & 8.5\\
\hline 
\end{tabular}
\end{center}
\vspace{-0.25in}
\end{table}
The contributions to the systematic error on the signal parameters are 
summarized in Table~\ref{tab:systematics}.
To estimate the errors due to the fit procedure, we perform fits on a large number of MC samples with  the proportions
of  signal, continuum and $B$-background events measured from data. 
Biases of a few percent
observed in these fits are due to imperfections in the likelihood model 
such as neglected correlations between the discriminating variables of the 
signal and $B$-background PDFs and are assigned as a systematic uncertainty of the fit procedure.  The error due to the  fit procedure includes the statistical error on the bias added in quadrature with the observed bias. 
The expected event yields from the $B$-background modes are varied according 
to the uncertainties in the measured or estimated branching fractions.  
Since $B$-background modes may exhibit  \CP violation, the corresponding 
parameters are varied within their physical ranges. 
We vary the parameters of the $\dt$ model and tagging fractions
 incoherently within their errors and assign as a systematic error 
the quadratic sum of the observed changes in our measured parameters.
The uncertainties 
due to the simulated signal PDFs are obtained from a control sample of 
fully reconstructed $B^{0} \rightarrow D^{-}(\to\KS\pi^-) \pi^{+}$ decays. 
The systematic errors  due to interference between the 
doubly-Cabibbo-suppressed  
(DCS) $\bar b \to \bar u c \bar d$ amplitude with the Cabibbo-favored 
$\bar b \to \bar c u \bar d$ amplitude for tag-side $B$ decays have been 
estimated from simulation by varying freely all relevant strong 
phases~\cite{bib:DCSD2003}.
The errors associated 
with $\dmd$ and $\tau$ are estimated by varying these parameters within
the errors on the world average~\cite{PDG2002}. 

The systematic error introduced 
in the Q2B approximation 
by ignoring interference effects between the $\fzeros$ and the other 
resonances  in the Dalitz plot (as listed earlier) is estimated 
from simulation by varying freely all relative strong phases and taking 
the largest observed change in each parameter as the error.  
While the systematic errors due to interference are comparable to the 
statistical error for the branching fraction and the $\fzeros$ mass 
and width, they are small compared to the  statistical error for $S$ and $C$.


\begin{figure}[t]
  \centerline{ 	\epsfysize3.0cm\epsffile{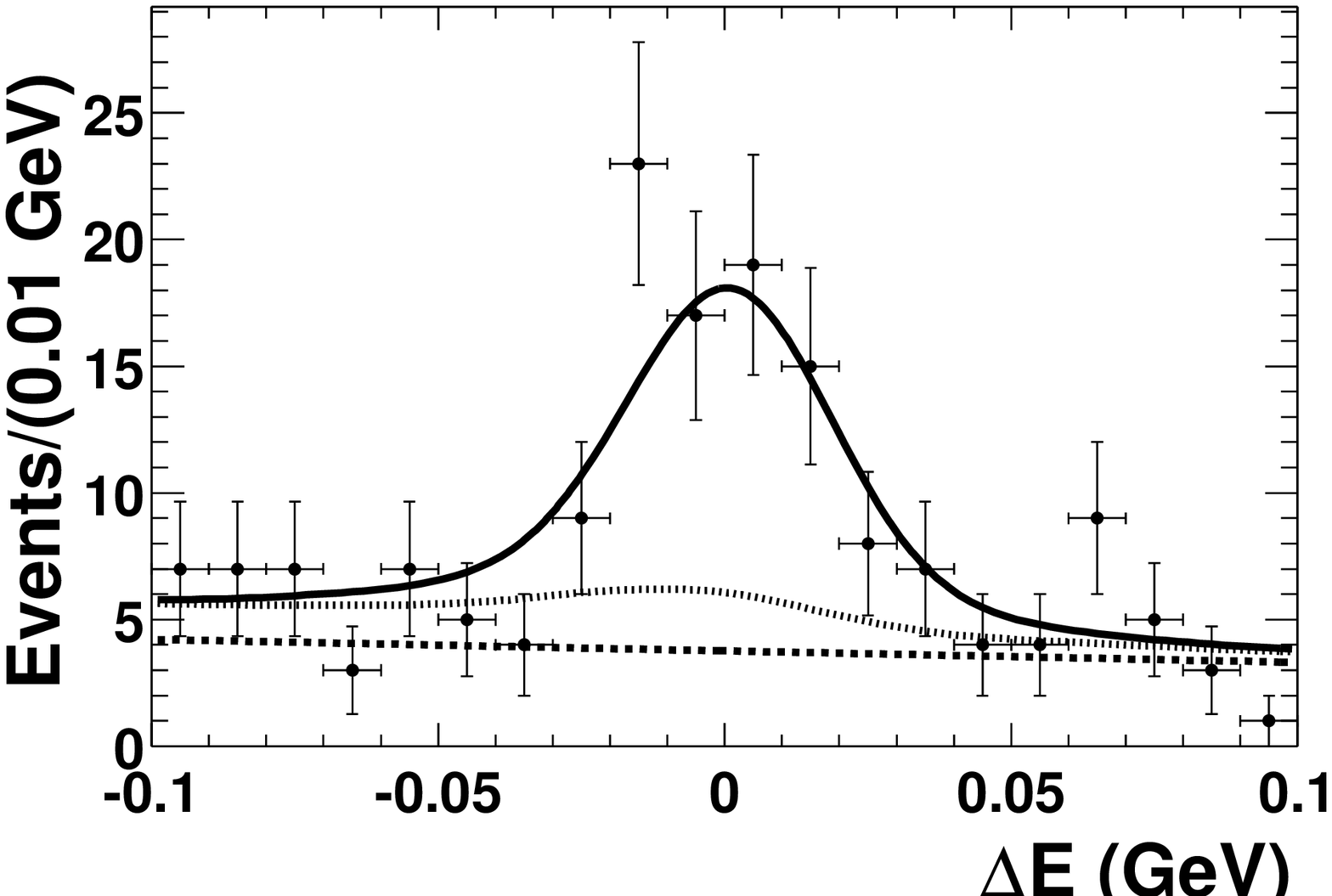}
		\epsfysize3.0cm\epsffile{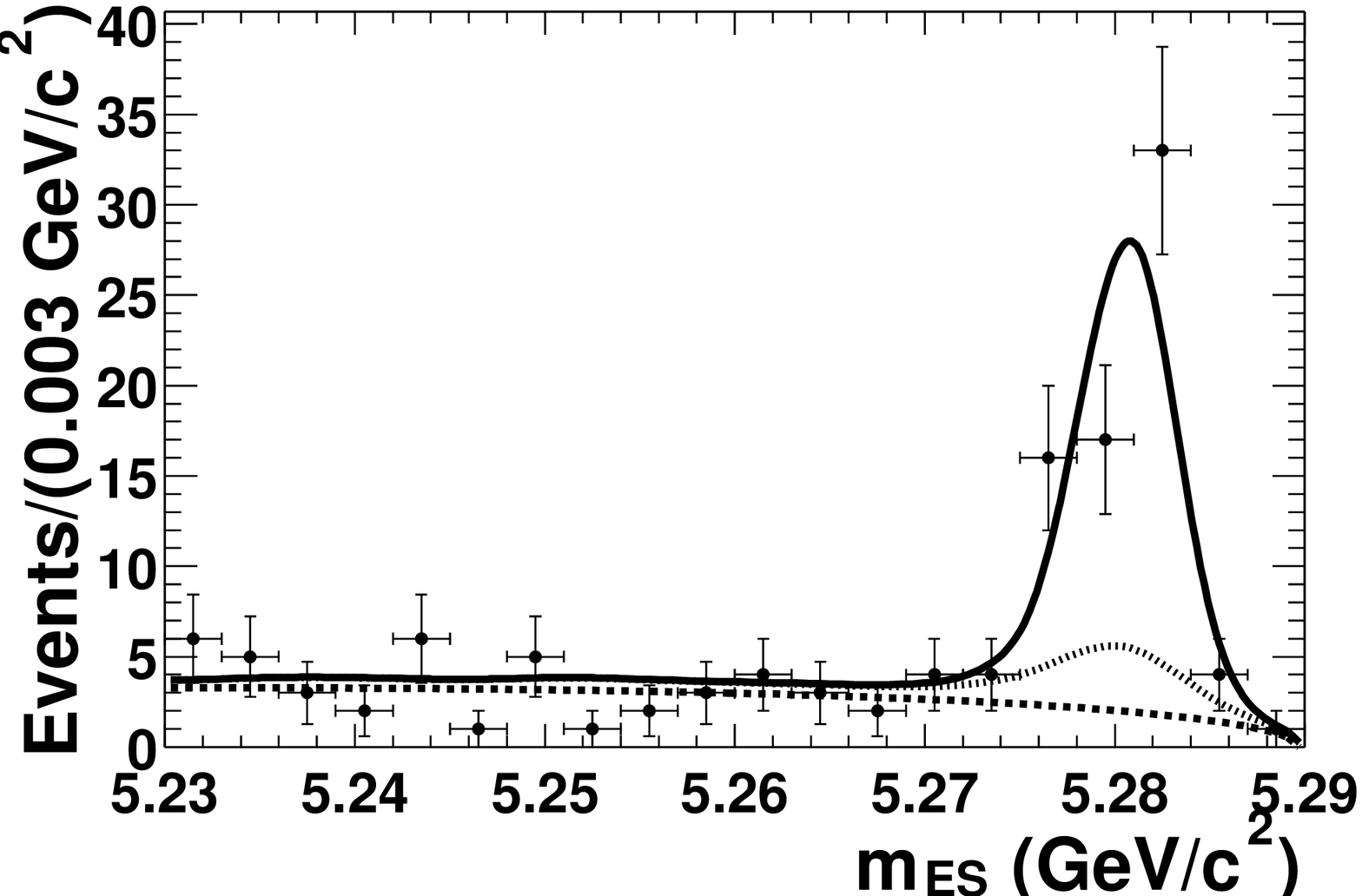}}
  \centerline{ 	\epsfysize3.0cm\epsffile{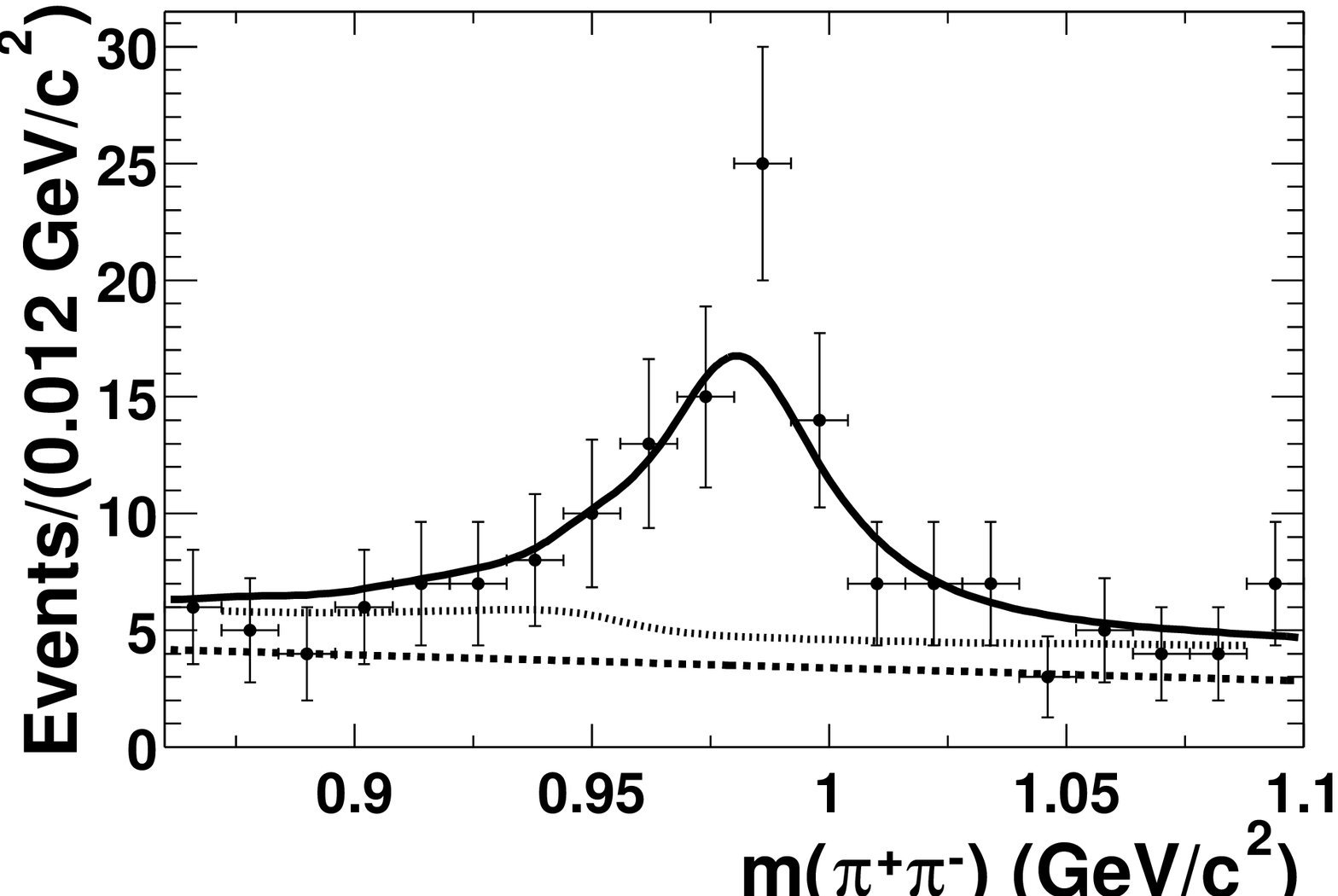}
                \epsfysize3.0cm\epsffile{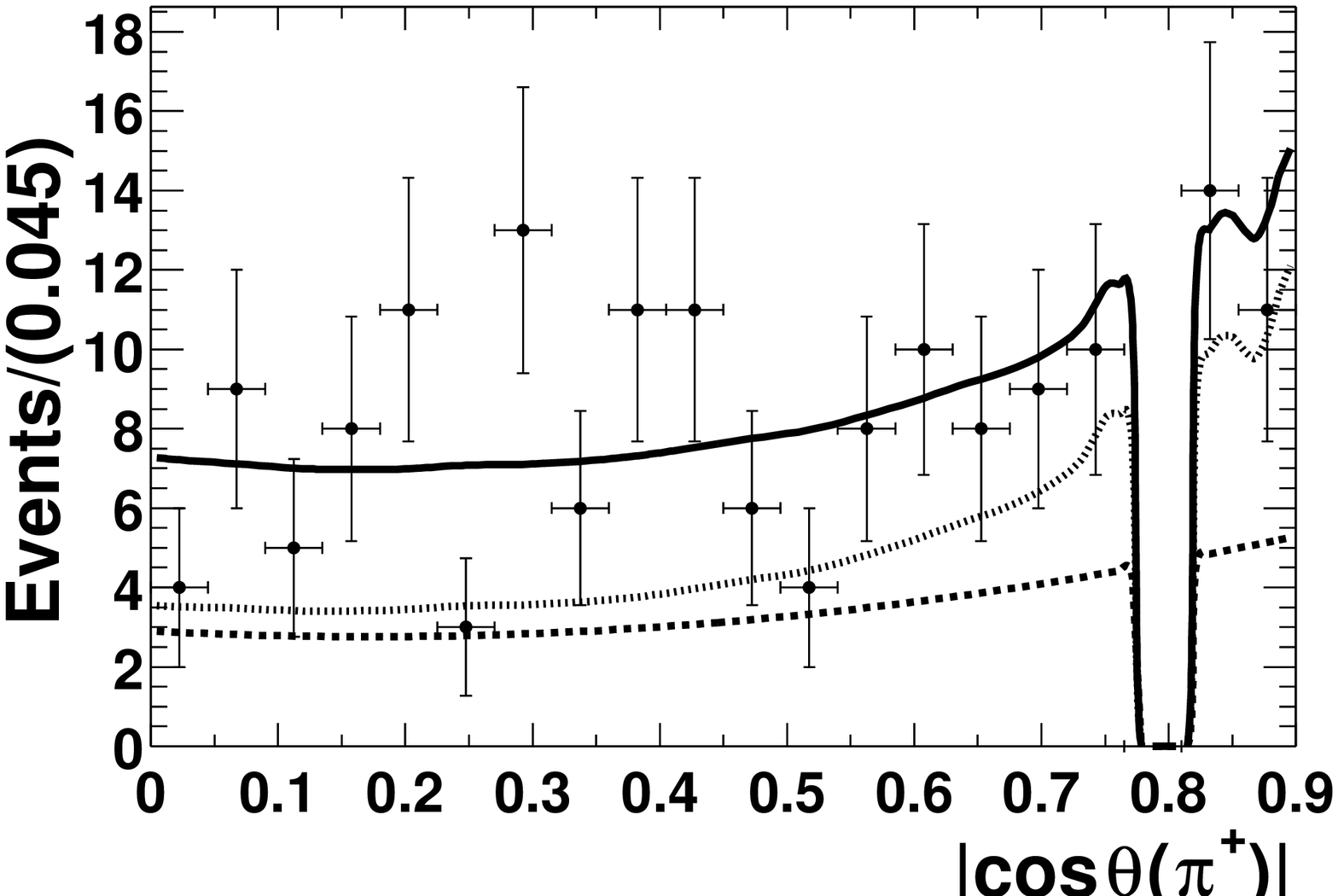}} 
\vspace{0.05cm}
\caption{\label{fig:ProjMesDE}
	Distributions of  (clockwise from top left) $\de$, $\mes$, 
	$|\cos{\theta(\pi^+)}|$, and $m(\pi^+\pi^-)$ for samples 
	enhanced in $\fzeros\KS$ signal.	The solid curve represents a 
	projection of the maximum likelihood fit result. The dashed 
	curve represents the contribution from continuum events, and 
	the dotted line (middle) indicates the combined contributions from 
	continuum events and $B$ backgrounds.  For presentation purposes, 
	the region  
	$0.765<|\cos\theta(\pi^+)|<0.81$ has been removed to suppress 
	the contribution from $D^-\pi^+$ events.}
\vspace{-0.20in}
\end{figure}
The maximum likelihood fit results in the $\fzeros\KS$ event yield 
$N_S=93.6\pm13.6\pm6.3$, where the first error is statistical and 
the second systematic.  The branching fraction corresponding to this 
yield is 
\begin{eqnarray*}
\lefteqn{
{\cal B}(\Bz\to\fzero K^0)\times{\cal B}(\fzero\to\pi^+\pi^-)\;=}\\
 && \hspace{2cm}
	(6.0 \pm 0.9 \pm 0.6\pm 1.2 )\tmsix\,,
\end{eqnarray*}
where the first error is statistical, the second systematic, and the 
third accounts for the model dependence in the quasi-two-body approximation.
The systematic error includes an uncertainty of $8.2\%$ from differences 
between data and MC in tracking, particle identification and $\KS$ 
detection efficiencies. Figure~\ref{fig:ProjMesDE} shows distributions 
of $\de$, $\mes$, $|\cos{\theta(\pi^+)}|$ and $m(\pi^+\pi^-)$,
that are enhanced in signal content by cuts on the signal-to-continuum 
likelihood ratios of the other discriminating variables. 
For the \CP-violation parameters, we obtain
\begin{figure}[t]
  \vspace{-0.4cm}
  \hspace{-0.2cm}
  \epsfysize9.0cm
  \epsffile{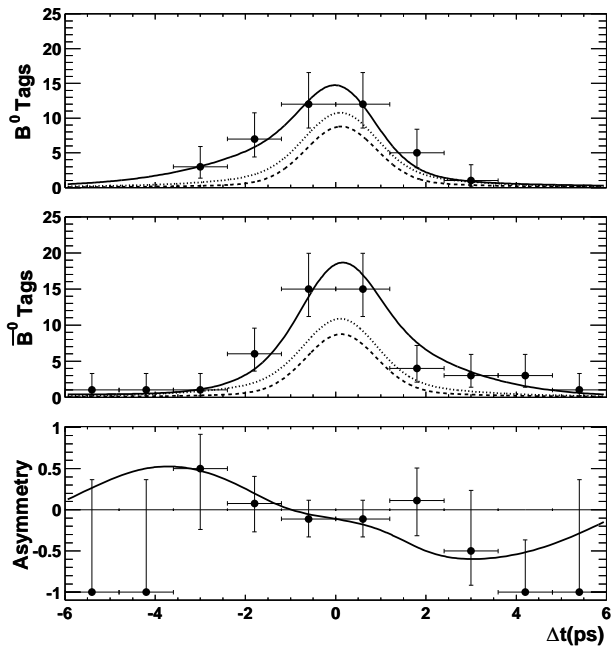}
\vspace{-0.3cm}
\caption{The signal enhanced time distributions tagged as
	 $\Bz_{\rm tag}$ (top) and  $\Bzb_{\rm tag}$ (middle), and the 
	asymmetry,  $A_{\Bz/\Bzb}$ (bottom). The solid curve
	is a projection of the fit result.
	The dashed line is the distribution for continuum background and the 
	dotted line is the total \B- and continuum-background 
	contribution. }
	\label{fig:asymCS} 
\vspace{-0.2in}
\end{figure}
\begin{eqnarray*}
	S	& = & -1.62^{\,+0.56}_{\,-0.51} \pm 0.09\pm 0.04\, ,\\
	C	& = &\phantom{-}0.27 \pm 0.36 \pm 0.10 \pm 0.07 \,.
\end{eqnarray*}
The  time-dependent distributions and  asymmetry
$A_{\Bz/\Bzb}=  (N_{\Bz} - N_{\Bzb})/(N_{\Bz} + N_{\Bzb})$ 
in the tagged events are represented in Fig.~\ref{fig:asymCS}.
The model-dependent mass and width 
of the $\fzeros$ are found to be
\begin{eqnarray*}
	m_{\fzeros}	& = & (980.6 \pm 4.1 \pm 0.5\pm 4.0)\mevcc \, , \\
	\Gamma_{\fzeros}	& = & (43 ^{\,+12}_{\,-9} \pm 3\pm9)\mevcc\, .
\end{eqnarray*}
These results are in agreement with previous mass and width 
measurements~\cite{PDG2002, E791Ds3pi}.

In summary, we have presented measurements of the branching fraction,
 resonance  parameters, and 
\CP-violating asymmetries in $\Bz\to\ffzeroppKs$ decays. 
Our results for $S$ and $C$ are consistent with the Standard Model 
at the $1.7\sigma$ and $0.8\sigma$ levels, respectively.
The result for $S$ is $1.2\sigma$ from the physical limit,
 and the hypothesis of no mixing-induced  \CP violation is
 excluded at the $2.7\sigma$ level.

We are grateful for the excellent luminosity and machine conditions
provided by our \pep2\ colleagues, 
and for the substantial dedicated effort from
the computing organizations that support \babar.
The collaborating institutions wish to thank 
SLAC for its support and kind hospitality. 
This work is supported by
DOE
and NSF (USA),
NSERC (Canada),
IHEP (China),
CEA and
CNRS-IN2P3
(France),
BMBF and DFG
(Germany),
INFN (Italy),
FOM (The Netherlands),
NFR (Norway),
MIST (Russia), and
PPARC (United Kingdom). 
Individuals have received support from CONACyT (Mexico), A.~P.~Sloan Foundation, 
Research Corporation,
and Alexander von Humboldt Foundation.



\begin{thebibliography}{99}
\bibitem{f0ref}  		Throughout the paper $\fzeros$ refers 
				to the $\fzero$ and its decay to $\pi^+\pi^-$.
				In addition, charge conjugate decay modes are 
				assumed unless explicity stated.
\end{thebibliography}
\end{document}